\documentclass{iau}
\usepackage{graphicx}

\title[The Road to Quasars] %% [give here short title] %%
{The Road to Quasars}

\author[K. I. Kellermann]   %% [give here the short author list; use "et al." if 3 authors or more] %%
{K. I. Kellermann}
\affiliation{
National radio Astronomy Observatory \\ 
520 Edgemont Rd, Charlottesville, VA 22903 \\ 
email: {kkellerm@nrao.edu} \\

}

\pubyear{2015}
\volume{313} 
\pagerange{xx--xx}
\jname{Extragalactic jets from every angle}
\editors{F. Massaro, C.C. Cheung, E. Lopez, A. Siemiginowska, eds.}
\begin{document}

\maketitle

\begin{abstract}
Although the extragalactic nature of 3C 48 and other quasi stellar radio sources was discussed as early as 1960 by John Bolton and others, it was rejected largely because of preconceived ideas about what appeared to be unrealistically high radio and optical luminosities.  Not until the 1962 occultations of the strong radio source 3C~273 at Parkes, which led Maarten Schmidt to identify 3C 273 with an apparent stellar object at a redshift of 0.16, was the true nature understood.  Successive radio and optical measurements quickly led to the identification of other quasars with increasingly large redshifts and the general, although for some decades not universal, acceptance of quasars as the very luminous nuclei of galaxies.
  
Curiously, 3C 273, which is one of the strongest extragalactic sources in the sky, was first cataloged in 1959 and the magnitude 13 optical counterpart was observed at least as early as 1887.  Since 1960, much fainter optical counterparts were being routinely identified using accurate radio interferometer positions which were measured primarily at the Caltech Owens Valley Radio Observatory.  However, 3C~273 eluded identification until the series of lunar occultation observations led by Cyril Hazard.   Although an accurate radio position had been obtained earlier with the OVRO interferometer, inexplicably 3C 273 was initially misidentified with a faint galaxy located about an arc minute away from the true quasar position.

\keywords{quasars, AGN.}
%% add here a maximum of 10 keywords, to be taken form the file <Keywords.txt>
\end{abstract}

\firstsection % if your document starts with a section,
              % remove some space above using this command.
\section{Introduction}

The discovery of radio galaxies and then quasars had a profound impact on development of astrophysics and cosmology during the later part of the 20th century.  Suddenly astronomers were faced with previously unimaginable luminosities coming from unimaginably small volumes.  Following the 1963 identification of 3C 273 at z=0.16, redshifts were quickly extended to z $\sim$ 2 and more.  But quasars could have, and indeed should have, been recognized earlier.  Indeed 3C 48 had been identified two years before 3C 273, but the apparent  redshift of 0.37 was rejected in favor of a galactic interpretation because of preconceived ideas about the limits to radio and optical luminosity.

Early suggestions about unusual activity in the nuclei of galaxies go back to the work of Sir James Jeans \cite{J61}.  But, the modern understanding of the important role of galactic nuclei probably began with the famous paper by Karl  
 \cite{Seyfert43}
who reported the presence of broad strong emission lines in the nuclei of seven spiral nebulae.  Interestingly, although Seyfert's name ultimately became attached to the general category of galaxies with broad nuclear emission  lines associated with highly ionized elements, his 1943 paper apparently went unnoticed until Baade and Minkowski \cite[(1954)]{BM54} drew attention to the similarity of the optical emission line spectrum of the galaxies studied by Seyfert with the galaxy that they had identified with the Cygnus A radio source. 
 
Earlier John Bolton, Gordon Stanley, and Bruce Slee 
\cite[(1949)]{BSS49} 
had identified three of the strongest discrete radio sources with the Crab Nebula, M87, and NGC 5128,  Until that time the discrete radio sources were widely thought to be associated with stars.  This was understandable, as Karl Jansky and Grote Reber had observed radio emission from the Milky Way.  The Milky Way is composed of stars, so it was natural to assume that the discrete radio sources had a stellar origin.  Bolton et al. understood the importance of the identification of the Taurus A radio source with the Crab Nebula which was widely recognized as the remnants of the 1054 supernova reported by Chinese observers. They correctly identified two other strong sources with M87 and NGC 5128, but realizing that their absolute radio luminosity would need to be $10^6$ times more luminous than that of the Crab Nebula, they  concluded that 

 \begin{quote}
\it{NGC 5128 and NGC 4486 (M87) have not been resolved into stars, so there is little direct evidence that they are true galaxies. If the identification of the radio sources are accepted, it would indicate that they are [within our own Galaxy]}. 
\end{quote}

They entitled their paper simply ``Positions of Three Discrete Radio Sources of Galactic Radio Frequency Radiation.''  John Bolton later explained that he really did understand that M87 and NGC 5128 were extraordinarily luminous extragalactic radio sources, but that he was concerned that that  a conservative Nature referee might hold up publication (personal communication from Bolton to the author, August, 1989).

The following years saw the identification of more radio galaxies, and the paradigm which had previously considered all discrete radio sources to be galactic stars quickly changed to one where to most high latitude sources were  assumed to be extragalactic.  The energy requirements were exacerbated with the identification of Cygnus A, the second strongest radio source in the sky with a magnitude 18 galaxy  which was at what was then considered a high redshift of 0.056 and a corresponding radio luminosity about $10^3$ times more luminous than M87 or NGC 5128 \cite[(Baade and Minkowski 1954)]{BM54}.  

By the end of the decade, many radio sources had been identified with optical counterparts, but the redshift of Cygnus A remained the largest of any known radio galaxy.  Typically the optical counterparts of strong radio sources were elliptical galaxies \cite{B60} that were the brightest member of a cluster.  In 1960, Rudolph Minkowski \cite[(1960)]{M60} identified 3C 295 with a magnitude 20 galaxy at z=0.46, by far the largest known redshift of any galaxy.  3C~295 is about ten times smaller than Cygnus A and ten times more distant, consistent with the idea that the smallest radio sources might be path finders to finding very distant galaxies. But a few months later Caltech radio astronomers identified the first of several very small radio sources with what appeared to be galactic stars, thus again raising questions about the extragalactic nature of other small diameter radio sources.

\section{The First Quasars}

In late 1960, while searching for ever more distant radio galaxies, Caltech radio astronomers, John Bolton and Tom Matthews identified 3C 48 with a faint magnitude 16 apparent stellar object.  Apparently Bolton suggested that 3C 48 might not be a galactic star but rather extragalactic with z=0.37.  But, shortly later Bolton left Caltech to take up his new responsibilities as the person in charge of completing the construction of the new 210-ft radio telescope at Parkes, which completely occupied him for the next few years.  At the 107th meting of the American Astronomical Society held in December, 1960, Allan Sandage \cite{S60} gave a late paper on "The First True Radio Star."

In an exhaustive unpublished  analysis of the complex 3C 48 emission line spectrum, Greenstein \cite{G62}  interpreted the observed emission lines in terms of highly ionized states of various rare earth elements.  He briefly speculated on a possible redshift of 0.37, but then dismissed the possibility that 3C 48 was extragalactic.   Nearly two years would pass, and two other compact radio sources, 3C 196 and 3C 286 were also identified as galactic stars.  Based on OVRO interferometer measurements which had a nominal position accuracy of  five or ten arcsec, in his 1962 PhD thesis, Caltech graduate student Dick  \cite{R62}  suggested an identification for 3C 273 with a faint galaxy,  which later turned to be located about an arc minute east of the currently accepted quasar position.  In May 1962, Maarten Schmidt tried to measure the position of Read's faint galaxy thought to be identified with 3C 273.  However, in his subsequent publication in the Astrophysical Journal submitted in December, 1962,  \cite{R63} dropped the suggested 3C 273 identification.    Although Read and other Caltech radio astronomers had measured the position of 3C 273 good to better than 10 arcsec, and the  existence of star-like optical counterparts was already well established, not until after the 1962 occultations at Parkes \cite{H63} showed the double source structure, would 3C 273 be correctly identified by \cite{S63} with a star-like object at a redshift of 0.16. Like 3C~48, the optical counterpart of 3C~273 contained a faint jet-like feature extending some 20 arcseconds away from the bright stellar object.

Interestingly, the first occultation positions  that were sent by John Bolton to Schmidt on August 20, just two weeks after the occultation, were in error by about 15 arcseconds. The correct radio positions were not known by Schmidt until after he received Bolton's letter of January 26, 1963, more than a month after Schmidt had observed the 3C 273 spectrum at Palomar.   Further details of the events surrounding the Parkes occultation observations and the subsequent interaction with Caltech and Palomar astronomers that led to the identification of the 3C 273 and the determination of its redshift are given in a companion paper in this volume by \cite{HJG15}, by Schmidt \cite{S83,S11} and by \cite{K13, K14}.

Applying the derived 3C~273 Balmer series redshift of 0.16, Schmidt also recognized the Mg II line at a rest wavelength of 2798 Angstroms, leading he and Greenstein  to determine that 3C 48 indeed was also extragalactic and had a  a redshift of 0.37 \cite{GM63}, just the value that was suggested two years earlier by Bolton, but rejected by Greenstein and Bowen.  The announcement of the 3C 48 redshift appeared as a companion paper to the Hazard et al. occultation position and the \cite{S63}  identification  with the quasar and its jet-like extension.  Although Schmidt's measurement and interpretation of the 3C 273 spectrum appears to have been critical in recognizing the extragalactic nature of 3C 48,  the 3C 48 paper appeared without Schmidt's or Bolton's names, but included Tom Matthews as an author, as Greenstein remembered that it was Matthews who first drew his attention to the optical counterpart \cite{G96}. 

In a 1989 lecture, John  \cite{B90} recalled that before leaving for Australia in November 1960, he had argued that 3C 48 had a redshift of 0.37 but that  Ira Bowen and Jesse Greenstein had convinced him that a 3 or 4 Angstrom discrepancy in the calculated rest wavelength of different lines was unacceptable.   Bolton's 1989 claim was immediately attacked  \cite{G96} as an attempt to rewrite history.  However, Bolton's assertion was substantiated by published memories of Fred \cite{H89}  and the much later discovery of a handwritten letter from Bolton to Joe Pawsey dated November 16, 1960 stating that 3C 48 had a redshift of 0.37.  However, a few weeks later when his ship stopped in Hawaii, Bolton again wrote to Pawsey saying "It's most likely a star."

\section{Radio Quiet Quasars}

The years following the recognition of the 3C 48 and 3C 273 redshifts led to the identification of more quasars at ever larger redshifts and the recognition that quasars are the extremely bright nuclei of active galaxies.  Generally, the identified quasars had a significant UV excess compared with stars, so due to the redshift of their spectrum, they appeared blue on photographic plates facilitating their identification with radio sources with even modest position accuracy \cite{RS64}.  

   In 1965, Sandage noted that the areal density of blue stellar objects was some thousand times greater than that of 3C radio sources.  Sandage argued that what he called quasi stellar galaxies were related to quasars, except that they were ``radio quiet.''  Sandage's paper was widely attacked, perhaps in part because it was received on May 15, 1965 at the Astrophysical Journal, but S. Chandrasekar, the ApJ  editor was apparently so impressed by Sandage's claim for a ``New Constituent of Universe'' that he held up publication of the Journal, so Sandage's paper would appear in the May 15 issue.  Tom Kinman \cite{K65} along with Lynds and Villere \cite{LV65} argued that most of Sandage's Blue Stellar Objects are just that, blue stellar objects, while Fritz Zwicky \cite{Z65} pointed out that he had previously called attention to this phenomena which he had called ``compact galaxies'' and he later accused Sandage of ``one of the most outstanding feats of plagerism'' \cite{Z71}.

As it turned out, most of Sandage's quasi stellar galaxies were indeed galactic and there are only some ten times more radio quiet quasars than radio loud quasars.   But, it has now been more than half a century since we have divided quasars into the two classes of radio loud and radio quiet quasars, and it is still has not been clear if there are two distinct populations or  whether the radio loud population is merely the extreme end of a continuous distribution of radio luminosity. Proponents of each interpretation claim that the other interpretation is due to selection effects.  Many of the previous investigations designed to distinguish between radio loud and radio quiet quasars were limited by contamination from low luminosity AGN with absolute optical magnitudes greater than -23, by the use of samples based on radio rather than optical selection criteria, and by inadequate sensitivity to detect radio emission from most of the radio quiet population.  

In an attempt to overcome these limitations, Kimball et al. \cite{K11} used the Jansky Very Large Array to observe 179 quasars selected from the SDSS. All of the  quasars were within the redshift range 0.2 to 0.3 and were brighter than $M_{i}$ = -23 so were legitimate quasars.  The radio observations were made at 6 GHs reaching an rms noise of $\sim$6 $\mu$Jy. All but about 6 quasars were detected as radio sources with an observed radio luminosity sharply peaked between $10^{22}$ and $10^{23}$ Watts/Hz characteristic of the radio luminosity typically observed from star forming galaxies.  About ten percent of the SDSS sample were found to be strong radio sources with radio luminosities ranging up to $10^{27}$ W/Hz.  Kimball et al.  concluded that the radio emission from radio quiet quasars is due to star formation in the host galaxy and is not directly related to the SMBH assumed to drive the large optical luminosity of quasars.

\section{Summary and Conclusions}

Quasars and AGN are now part of the normal lexicon of astronomy. Although the 1962 occultations of 3C 273 were crucial in bringing 3C 273 to the attention of Maarten Schmidt, the occultation position used by Schmidt for his Palomar observations was in error by about 15 seconds of arc.  A more accurate position had been established from OVRO interferometer observations but apparently was not accurately conveyed to Schmidt.

Prior to the Parkes occulation, 3C 273 was not on anyone's list of potentially interesting sources to be further studied, as it was already known not to be very small and thus presumably (and correctly) not very distant.  Schmidt's December 1962  spectroscopic observations  of 3C 273 were made  only because he had received the occultation position from Bolton.  However, since the correct radio position was not available until after Schmidt had recognized that the optical counterpart was not galactic, but had a redshift, z=0.16, the occultation positions did not directly result in the identification of the optical counterparts to the radio positions. Rather, not having an accurate radio position, Schmidt \cite{S83,S11}  believing that the correct optical counterpart was the jet, apparently took a spectrum of the quasar itself only to ``get it out of the way.''  Ironically, it was possible for Schmidt to recognize the ``large'' redshift of 3C 273, because, unlike for 3C 48, the redshift of 3C 273 was ``small,'' and the familiar Balmer spectrum was not shifted out of the optical window.

The earlier identification of 3C 48, and later the two other apparently stellar counterparts of 3C 196 and 3C 286 played no role in the recognition of quasars.  The 1960 rejection by Greenstein, Bowen, and ultimately also by Bolton of the suggested 3C 48 redshift  was explained by a 3 to 4 angstrom discrepancy in the derived rest wavelengths, an amount that might easily have been understood as the result of Doppler motions, but more likely was their unwillingness to accept the implications of the extraordinary luminosity  and small dimensions implied by the optical variability.  

Following the recognition of the 3C 48 and 3C 273 redshifts, new identifications quickly led to redshifts  up to z $\sim$2.  But, contrary to hopes, because they are not standard candles, quasars have had little impact to classical cosmology and the quest to define $H_0$ or $q_0$.  However, discovery of quasars has led to the recognition that they are driven by a central engine, widely believed to be due to accretion onto a SMBH of $\sim 10^9$ or more solar masses.

Over the half century since their discovery, quasars have had a profound impact to the sociology of astronomy and astronomers as well as to astrophysics and cosmology.  For the first decade or more, there was vocal minority that continued to advocate a non cosmological interpretation of quasar redshifts.  The controversy was intense and sometimes very personal.  The arguments never died, until their proponents died.

\section{Unanswered Questions}

$\bullet$  Why did  Greenstein, Matthews, Sandage and others wait two years to submit papers interpreting  3C 48 as a galactic star. Why did John Bolton wait nearly thirty years before going public with his claim to the discovery of the 3C 48 redshift?

$\bullet$  Why did it take so long to identify 3C 273 with such a bright magnitude 13 optical counterpart, even though by the time of the 1962 Parkes occultation, at least three other sources, 3C 48, 3C 147, and 3C 196 had already been identified with apparent stellar counterparts, and a Caltech interferometer position was apparently known to an accuracy of 5 or 10 arcseconds  by the time of Schmidt's May 1962 Palomar observations.

$\bullet$ Why was the extragalactic nature of 3C 48 not accepted earlier as suggested by John Bolton in November, 1960 and considered by Greenstein in his unpublished manuscript submitted to the Astrophysical Journal?   Why was a 3 or 4 Angstrom discrepancy among the derived rest wavelengths considered fatal to the z=0.37 interpretation when just a few months earlier Minkowski had gone out on a limb with the z=0.46 redshift based on only one spectral line and a comparable radio luminosity?  Why were Bolton and Schmidt ignored in the 1963 publications by Greenstein and Matthews (1963) and by \cite{MS63}? 

$\bullet$  Why did it take Maarten Schmidt six weeks to decipher the simple Balmer spectrum of 3C 273?  Was it possibly due to the intervention of  holiday activities?  Or, perhaps, it was the receipt of Bolton's January 26 letter, with improved radio positions, that inspired Schmidt to re-inspect and finally understand the 3C 273 spectrum six weeks after he had observed it. 

$\bullet$  Although all quasars presumably contain a SMBH which drives the excessive luminosity, why are only about 10\% of optically selected quasars also strong radio sources?  
  
\section{Acknowledgment}  The National Radio Astronomy Observatory is operated by Associated Universities Inc. under Cooperative Agreement with the National Science Foundation.   I am grateful to many colleagues, especially Maarten Schmidt, Tom Matthews, Jesse Greenstein, Allan Sandage, Cyril Hazard, and John Bolton who have shared with me their recollections of the events surrounding the discovery of quasars.   Miller Goss and Ron Ekers brought the 1960 and 1961 letters from John Bolton to Joe Pawsey to my attention.  Dave Jauncey provided valuable comments, and John Faulkner alerted me to the 1981 recollection by Hoyle of Bolton's 1960 claim that 3C 48 was extragalactic with z=0.37.

\end{document}